\documentclass[aps,apl,superscriptaddress,showpacs,reprint,floatfix]{revtex4-2}
\usepackage[hidelinks]{hyperref}
\usepackage{graphicx}
\usepackage{gensymb}
\usepackage{amsmath}
\usepackage{amssymb}
\usepackage{physics}
\usepackage{mathtools}
\usepackage{graphicx}
\usepackage{bm}
\usepackage{xcolor}
\usepackage{enumitem}
\usepackage[normalem]{ulem}

\newcommand{\Ca}{\ensuremath{^{40}{\rm Ca}^+\,}}

\begin{document}

\title{Probing rotational decoherence with a trapped-ion planar rotor}

\author{Neil Glikin}
\affiliation{Department of Physics, University of California, Berkeley, Berkeley, California 94720, USA}
\affiliation{Challenge Institute for Quantum Computation, University of California, Berkeley, Berkeley, California 94720, USA}

\author{Benjamin A. Stickler}
\affiliation{Institute for Complex Quantum Systems and Center for Integrated Quantum Science and Technology, Ulm University, Albert-Einstein-Allee 11, 89069 Ulm, Germany}

\author{Ryan Tollefsen}
\affiliation{Department of Physics, University of California, Berkeley, Berkeley, California 94720, USA}
\affiliation{Challenge Institute for Quantum Computation, University of California, Berkeley, Berkeley, California 94720, USA}

\author{Sara Mouradian}
\affiliation{Department of Electrical and Computer Engineering, University of Washington, Seattle, Washington 98195, USA}

\author{Neha Yadav}
\affiliation{Department of Physics, University of California, Berkeley, Berkeley, California 94720, USA}
\affiliation{Challenge Institute for Quantum Computation, University of California, Berkeley, Berkeley, California 94720, USA}

\author{Erik Urban}
\thanks{Current address: Exponent, Inc., Electrical Engineering and Computer Science Practice, Warrenville, IL}
\affiliation{Department of Physics, University of California, Berkeley, Berkeley, California 94720, USA}

\author{Klaus Hornberger}
\affiliation{University of Duisburg-Essen, Faculty of Physics, Lotharstra\ss e 1, 47057 Duisburg, Germany}

\author{Hartmut H\"affner}
\affiliation{Department of Physics, University of California, Berkeley, Berkeley, California 94720, USA}
\affiliation{Challenge Institute for Quantum Computation, University of California, Berkeley, Berkeley, California 94720, USA}

\begin{abstract}
The quantum rotor is one of the simplest model systems in quantum mechanics, but only in recent years has theoretical work revealed general fundamental scaling laws for its decoherence. For example, a superposition of orientations decoheres at a rate proportional to the sine squared of the angle between them. Here we observe scaling laws for rotational decoherence dynamics for the first time, using a 4\,\textmu m-diameter planar rotor composed of two Paul-trapped ions. We prepare the rotational motion of the ion crystal into superpositions of angular momentum with well-defined differences ranging from $1-3\,\hbar$, and measure the rate of decoherence. We also tune the system-environment interaction strength by introducing resonant electric field noise. The observed scaling relationships for decoherence are in excellent agreement with recent theoretical work, and are directly relevant to the growing development of rotor-based quantum applications.
\end{abstract}

\maketitle

\emph{Introduction.}---The quantum rigid rotor is among the simplest model quantum systems, on par with the quantum harmonic oscillator and the qubit. In contrast to these, the technological potential of quantum rotor dynamics has not yet been fully tapped. There have recently been remarkable advances in the preparation and control of molecular rotational states \cite{Koch2019, Chou2020, Lin2020, Sinhal2020, Voges2020, Schindewolf2022, Stevenson2023}, in using such states for coherent interactions \cite{Yan2013, Bao2023, Christakis2023, Li2023, Gregory2024, Holland2023}, and in the manipulation of nanoscale rotors \cite{Delord2020, Ahn2020, VanDerLaan2021, Stickler2021, Pontin2023, Kamba2023}. These developments are motivated by prospects of using the unique properties of rotors for encoding quantum information \cite{DeMille2002, Grimsmo2020, Albert2020, Jain2024, Furey2024}, performing quantum simulations using tunable dipole-dipole interactions \cite{Zhou2011, Lechner2013, Sundar2018, Guo2023}, torque sensing \cite{Ahn2020, Vinante2020, Ju2023}, and tests of the quantum superposition principle at high masses \cite{Stickler2018, Ma2020}. 

Experiments which aim to exploit rotational quantum coherence will also need to contend with interaction with their environment and the resulting decoherence. In rotor systems, decoherence has been experimentally observed only in certain limited contexts. These include the decay of rotational coherences in gas-phase ensembles of molecules due to collisions or radiation \cite{Vieillard2008, Owschimikow2010, Milner2014, Tenney2016, Damari2017}, and due to light scattering or uncontrolled differential shifts in optical traps \cite{Neyenhuis2012, Yan2013, Seesselberg2018, Bause2020, Burchesky2021, Gregory2024}. To push quantum rotors into a useful coherent regime, a comprehensive understanding of their decoherence dynamics will be crucial.

The theory of rotor decoherence, which fully accounts for orientational periodicity, has been developed only within the last decade \cite{Stickler2016a, Zhong2016, Papendell2017, Stickler2018b, Schlosshauer2019}. An important case is interaction with an isotropic environment, which leads to diffusion of a rotor's angular momentum state. In this case, a superposition of two orientation states $\ket{\phi}+\ket{\phi'}$ is predicted to decay into a statistical mixture at a rate proportional to two important parameters: (i) the angular momentum diffusion coefficient, and (ii) for inversion-symmetric planar rotors, the sine squared of their relative angle, $\sin^2(\phi-\phi')$. This can be compared with the decoherence rate $\abs{\alpha-\alpha'}^2$ for a superposition of harmonic oscillator coherent states $\ket{\alpha}+\ket{\alpha'}$ \cite{Walls1985}. Experimentally verifying harmonic oscillator decoherence scalings \cite{Brune1996, Myatt2000, Auffeves2003} proved crucial for developing oscillators into building blocks for quantum technologies. For rotors, the analogous relations have not yet been experimentally probed.

Here we report on an experimental study of the decoherence dynamics of a planar rigid rotor. The rotor is composed of two trapped atomic ions separated by several micrometers and rotating with a frequency on the order of one hundred kilohertz. We prepare this system in a superposition of different rotation rates, which turns into a superposition of orientation states, and engineer a noisy electric-field environment to induce rotational decoherence on the millisecond timescale. Our results demonstrate a sine-squared scaling of the decoherence rate for orientational superpositions in the small-angle limit and the expected scaling with the angular momentum diffusion coefficient. Both findings are in excellent agreement with theory.

\begin{figure*}
    \centering
    \includegraphics[width=\textwidth]{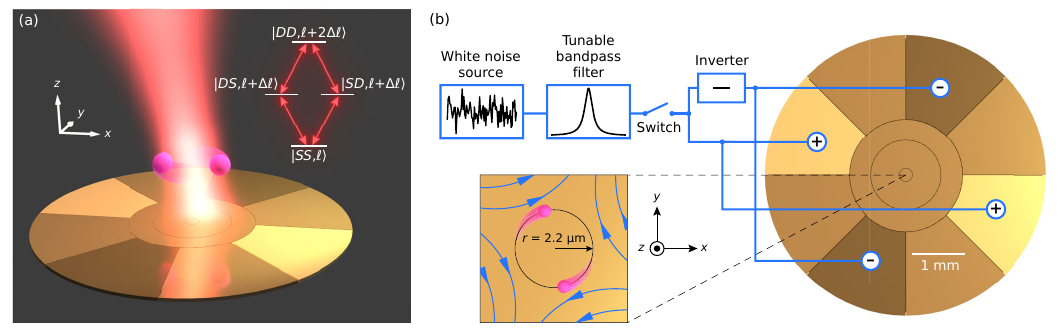}
    \caption{Experimental setup.
    (a) Two \Ca ions (not to scale) rotate above a circular surface ion trap in the plane parallel to the surface. A tilted vertical beam of 729\,nm laser light addresses the ion crystal on rotational sidebands of the electronic $\ket{S}\leftrightarrow\ket{D}$ transition, labeled by their order $\Delta\ell$, to create angular momentum superpositions. Sidebands of different orders $\Delta\ell$ are spectrally resolved from one another.
    (b) Circuit schematic of the setup for creating the electric-field gradient noise environment. A narrow-band noisy voltage is applied to the outer control electrodes in an alternating pattern, giving rise to a quadrupole electric field at the location of the ion rotor. The zoom-in shows the ion rotor and the quadrupole field (blue). The noisy gradient of this field exerts random torques on the rotor when on resonance with the ions' rotation, resulting in angular momentum diffusion and decoherence.}
    \label{fig:setup}
\end{figure*}

\emph{Two-ion planar rotor.}---We realize a quantum rigid rotor as a pair of \Ca ions Paul-trapped in a cylindrically symmetric potential, created by a surface ion trap with circular radio-frequency (rf) electrodes, see Fig.~\ref{fig:setup}. The ions are confined to the tranverse plane, while the cylindrical symmetry allows for a free rotational motion of the ion crystal within the plane~\cite{Li2017, Urban2019}. The ions' mutual Coulomb repulsion balanced with the trap potential creates a rigid rotor with radius $r = 2.2$\,\textmu m. The Hamiltonian describing the rotational motion of the ion crystal is that of a planar rotor $H_0 = L_z^2/2I$, where $L_z$ is the angular momentum operator component normal to the rotor plane and $I = 2mr^2$ is the moment of inertia. The angular momentum eigenstates $\ket{\ell}$ are simultaneous eigenstates of the Hamiltonian $H_0$ and the angular momentum operator $L_z$, with angular momentum eigenvalues $\hbar \ell $ where $\ell$ spans the integers. In the angle basis $\ket{\phi}$ they take the form $\bra{\phi}\ell\rangle = \exp(i \ell \phi)/\sqrt{2\pi}$. The energy scale is set by the rotational constant $\omega_{\rm r} = \hbar/2I = 2\pi\times13$\,Hz. Each ion contains an independent electronic degree of freedom, of which only two states are of interest: the $^2S_{1/2}\, (m=-1/2)$ sublevel of the ground electronic state (labelled $\ket{S}$) and the metastable $^2D_{5/2}\, (m=-1/2)$ state (labelled $\ket{D}$). The ions' internal electronic states are coherently manipulated by a narrow-linewidth laser near the $\ket{S}\leftrightarrow\ket{D}$ resonance at 729\,nm. Sidebands of this resonance are used to couple to the rotational degree of freedom. All measurements record the $\ket{D}$ state population averaged between the two ions, $P(D)$.

\emph{Controlled environment.}---We experimentally realize a controllable environment by applying a noisy voltage to the outer control electrodes of the trap, giving rise to a noisy electric field that couples to the rotor via the ions' charge. Since the two-ion crystal has zero dipole moment, the interaction with the noise field is via its quadrupole component and thus is proportional to the local field gradients,
\begin{equation}\label{randompotential}
    V_t(\phi) = \frac{er^2}{4} \left (\varepsilon_t L_+^2 + \varepsilon_t^* L_-^2\right ).
\end{equation}
Here, $L_\pm = e^{\pm i \phi}$ are dimensionless angular momentum raising and lowering operators, $L_\pm \ket{\ell} = \ket{\ell \pm 1}$ \cite{Edmonds1996} and $\varepsilon_t = -\partial_x E_x + \partial_y E_y + 2i \partial_x E_y$ is the relevant field gradient for quadrupolar interaction in the plane parallel to the surface \cite{supp}.  We inject voltage noise onto four of the eight control electrodes which surround the rf electrodes with alternating sign, as shown in Fig.~\ref{fig:setup}(b). The resulting quadrupole field is proportional to the applied voltage, which is generated as white noise from an arbitrary waveform generator and passed through a bandpass filter with a 19\,kHz bandwidth and tunable center frequency. This allows us to directly control the noise spectral density of the electric field gradient. In the absence of injected voltage noise, uncontrolled ambient electric field gradient noise is still present at the location of the ions due to stray voltage noise and thermally induced field fluctuations originating from the electrodes \cite{Brownnutt2015, Martinetz2022}.

Averaging the rotor-field interaction over many realizations of the noisy electric field gradient, the coarse-grained dynamics of the rotor's quantum state $\rho$ are described by the Lindblad master equation of orientational decoherence for quadrupolar interactions $\partial_t \rho = -i [H_0,\rho]/\hbar + \mathcal{L}\rho$ with \cite{supp},
\begin{equation}
    \mathcal{L}\rho = \frac{D}{4\hbar^2} \left(L_-^2\rho L_+^2 +  L_+^2\rho L_-^2 - 2\rho\right).
    \label{eq:master-equation}
\end{equation}
The Markovian description is only valid for timescales large in comparison to the inverse linewidth of the noise spectrum. The master equation describes angular momentum diffusion with a coefficient $D$, i.e. $\partial_t \expval{L_z^2} = 2 D$, and decoherence in the rotor orientation \cite{Papendell2017},
\begin{equation}
    \bra{\phi}\mathcal{L}\rho\ket{\phi'} = -\frac{D}{\hbar^2} \sin^2(\phi-\phi')\, \bra{\phi}\rho\ket{\phi'}.
    \label{eq:orientational-decoherence-rate}
\end{equation}
The scaling with $\sin^2(\phi-\phi')$ is $\pi$-periodic rather than $2\pi$-periodic due to the inversion symmetry of the quadrupole interaction \cite{Papendell2017, Pedernales2020}. For a rapidly spinning rotor with mean angular frequency $\omega_{\rm rot}=\expval{L_z}/I$, the diffusion coefficient is determined by the power spectral density of the field gradient at twice the rotation frequency \cite{supp}, $D = e^2r^4S_{\varepsilon}(2\omega_\text{rot})/4$. Here, $S_\varepsilon(\omega)$ is the power spectral density of $\varepsilon_t$. Equation~\eqref{eq:orientational-decoherence-rate} describes how angular momentum diffusion, which can also be understood via the classical interaction between the rotor system and the electric-field gradient environment, relates to quantum decoherence of the rotor.

\emph{Rotational diffusion and decoherence.}---We present two types of measurements, (i) angular momentum diffusion of the quantum rotor, and (ii) decoherence of angular momentum superpositions. In both cases, we begin with the ions trapped in an pinning potential which fixes their locations. We first cool all motional degrees of freedom with Doppler cooling, then sideband cool the transverse librational motion and the motion normal to the rotor plane to their vibrational ground states. We then initialize the ions' internal states to $\ket{S}$ via optical pumping, and spin up the rotor by rotating and then releasing the pinning potential via a protocol described in Ref.~\cite{Urban2019}. This approximately produces a coherent rotor state $\ket{\Psi_{\overline{\ell}}}$ that is localized in angular momentum space at $\bar{\ell} = \expval{L_z}/\hbar  \gg \sigma_\ell$, where $\sigma_\ell$ is the standard deviation in units of $\hbar$. In the present experiment, $\bar{\ell} \approx 6\times10^3$ and $\sigma_\ell \approx 20$. The corresponding mean rotation frequency, $\omega_\text{rot} = 2\omega_{\rm r}\bar{\ell}$, is chosen to be between 142 and 149\,kHz. This procedure yields resolved sidebands on the $\ket{S} \leftrightarrow \ket{D}$ 729\,nm transition at frequencies $\Delta\ell\,\omega_\text{rot}$ for integer $\Delta\ell$, corresponding to rotational state transitions $\ket{\ell} \to \ket{\ell+\Delta\ell}$. This allows selection of well-defined angular momentum changes $\Delta\ell$ by tuning the frequency of the 729\,nm laser to the corresponding sideband frequency.

\begin{figure}
    \centering
    \includegraphics[width=0.49\textwidth]{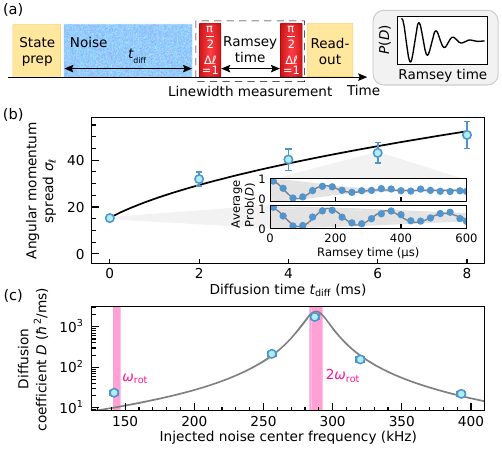}
    \caption{Measurements of the angular momentum diffusion coefficient.
    (a) Diagram of the measurement protocol for the angular momentum spread. A rotating coherent state is prepared, and noise is injected for a time $t_{\rm diff}$. Afterward, the linewidth of the $\Delta\ell=1$ sideband is measured with a Ramsey experiment. The decay time of the Ramsey signal is inversely proportional to the sideband linewidth, which is directly proportional to the angular momentum spread of the rotor.
    (b) A sample diffusion coefficient measurement. Angular momentum distribution width $\sigma_\ell$ is measured as a function of time after rotational state preparation. The solid line is a fit giving $D = 156(20)\,\hbar^2$/ms. Inset: Selected individual Ramsey measurements of $\sigma_\ell$ with a detuning of 6\,kHz from the $\Delta \ell = 1$ sideband transition. The shaded region shows the inferred phase contrast from the data, from  which $\sigma_\ell$ is inferred.
    (c) Demonstration of the $2\omega_\text{rot}$ resonance at a rotation frequency $\omega_\text{rot} = 144(2)$\,kHz. For each data point, the center frequency of the injected noise spectrum is shifted, while the amplitude and bandwidth are preserved, and the rotation frequency is kept constant. Measured diffusion coefficients below 20\,$\hbar^2$/ms are likely prevented here by ambient noise. A Lorentzian lineshape, centered at $2\omega_\text{rot}$ and with bandwidth equal to the injected noise bandwidth of 19\,kHz, is shown for reference. Vertical bands are at $\omega_\text{rot}$ and $2\omega_\text{rot}$ including variation.}
    \label{fig:diffusion-measurements}
\end{figure}

To determine the angular momentum diffusion coefficient, we measure the width of the rotor state $\sigma_\ell$ over time as it interacts with the environment. Under diffusion, this width increases with time as $\sigma_\ell(t_\text{diff})^2 = \sigma_\ell(0)^2 + 2Dt_\text{diff}/\hbar^2$. The measurement protocol is shown in Fig.~\ref{fig:diffusion-measurements}(a). After letting the rotor interact with the environment for time $t_{\rm diff}$, we perform a Ramsey experiment on the $\Delta\ell=1$ sideband to measure its linewidth, which is equal to $4\omega_{\rm r}\sigma_\ell$ \cite{Urban2019}. The linewidth is directly proportional to $\sigma_\ell$ since uncertainty in angular momentum directly corresponds to uncertainty in rotation frequency. We add a small detuning of a few kHz to see Ramsey fringes. The linewidth is inferred to be the reciprocal of the fringe decay time \cite{Urban2019}. We find that in the absence of injected noise, the uncontrolled ambient environment induces diffusion coefficients between 2 and 20 $\hbar^2$/ms.

Figure~\ref{fig:diffusion-measurements}(b) shows a sample diffusion measurement. In Fig.~\ref{fig:diffusion-measurements}(c), we plot the measured diffusion coefficient for noise spectral densities with varying center frequency. We find that the measured diffusion coefficient is indeed proportional to the field gradient spectral density: diffusion is maximized when the injected noise resonance is equal to $2\omega_\text{rot}$, and is consistent with a Lorentzian lineshape with bandwidth of 19\,kHz as the center frequency of the noise is swept. 

\begin{figure*}
    \centering
    \includegraphics[width=\textwidth]{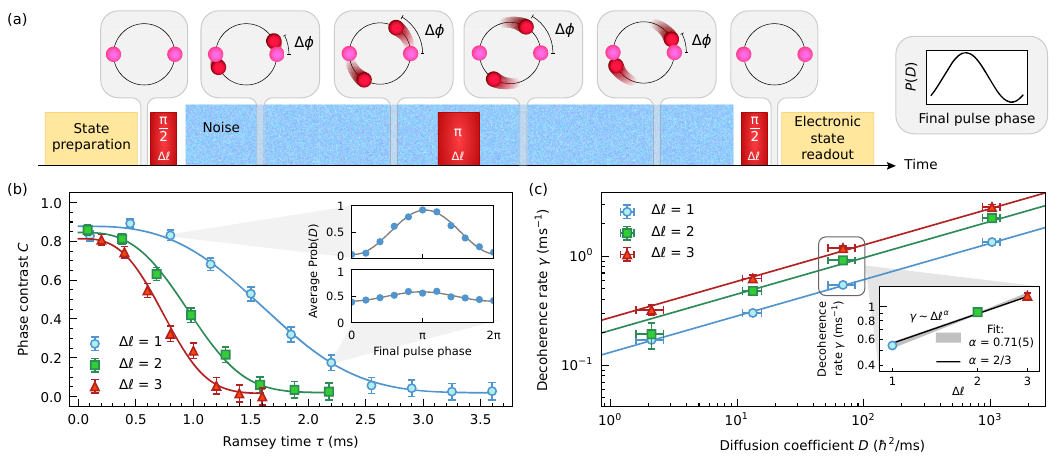}
    \caption{Measurements of rotational decoherence.
    (a) Simplified diagram of the Ramsey sequence used to measure rotational decoherence. Ions are shown in a co-rotating frame. A $\pi/2$ pulse on a rotational sideband of order $\Delta\ell$ creates a superposition of angular momenta. This may be equated to a superposition of two orientations of the rotor which are freely evolving at different angular velocities and have a linearly increasing angular separation $\Delta\phi$. A $\pi$ pulse reverses the separation, and a $\pi/2$ pulse closes the interferometer. The noisy environment is turned on while the interferometer is open. The remaining orientational coherence is quantified as the amplitude of the interference fringe. 
    (b) Sample measurements of the decay of phase contrast for superposition states with $\Delta\ell = $1, 2, 3 at a constant level of noise injected such that $D = 70\,\hbar^2$/ms. Fits are of the form \eqref{eq:Ct-highD}. Inset: Selected individual phase scans for $\Delta\ell=1$. 
    (c) Decoherence rate measured against angular momentum diffusion coefficient. The leftmost datapoints are derived from an ambient environment where $D = 2.1\,\hbar^2$/ms, and the others are obtained from use of the engineered environment, with varying amplitudes of injected noise. Solid curves are produced from \eqref{eq:gamma}, with no free fitting parameters. Inset: Decoherence rates, from the data points at $D=70\,\hbar^2$/ms in the main plot, plotted against $\Delta\ell$ on a log-log scale. The shaded area shows a power law fit including uncertainty. The black line shows the theoretically expected power law of $\Delta \ell^{2/3}$ from Eq.\,\eqref{eq:gamma}.}
    \label{fig:decoherence-measurements}
\end{figure*}

A direct coherence measurement is required to verify that the even classically expected angular momentum diffusion is accompanied by decoherence, which can only be understood in a quantum picture. For this, we perform a Ramsey interference experiment on a chosen $\Delta\ell$ sideband in the presence of the noise bath. The initial $\pi/2$ pulse applies the operation $\ket{SS,\ell} \to \left(\ket{SS,\ell} \pm \ket{SD,\ell\!+\!\Delta\ell} - \ket{DS,\ell\!+\!\Delta\ell} \mp \ket{DD,\ell\!+\!2\Delta\ell}\right)/2$, where the signs depend on whether $\Delta\ell$ is odd (top) or even (bottom). Because the initial state consists of a spread of many angular momentum values $\ell$, each of which has a different $\ket{\ell}\to\ket{\ell+\Delta\ell}$ transition frequency, free coherent evolution following the $\pi/2$ pulse results in a dephasing when tracing over the motion \cite{Urban2019}. We therefore add a Hahn echo pulse to the center of the Ramsey scheme to reverse the free evolution and in turn compensate for the dephasing. After time $\tau$, the interferometer is closed by a final $\pi/2$ pulse. With the echo pulse included, any measured loss in phase contrast can be attributed to nonunitary dynamics. We rule out other known potential decoherence mechanisms which could otherwise contribute to decay in fringe contrast by independent measurements \cite{supp}. The angular momentum superposition created by this experiment is akin to a time-dependent superposition of orientations which separate at a rate $2\omega_{\rm r}\Delta\ell$, as depicted in Fig.~\ref{fig:decoherence-measurements}(a).

A sample of rotational decoherence measurements is presented in Fig.~\ref{fig:decoherence-measurements}(b), which shows the Ramsey phase contrast $C$ as a function of total Ramsey time $\tau$ for angular momentum superposition orders $\Delta\ell = $ 1, 2, and 3. The expected fringe contrast can be calculated by solving the master equation \eqref{eq:master-equation} \cite{supp}, yielding
\begin{equation}
    C(\tau) = \exp\left(-\frac{D\tau}{2\hbar^2} \big[1-\text{sinc}(2\omega_{\rm r}\Delta \ell\, \tau)\big]\right),
    \label{eq:Ct-general}
\end{equation}
which describes an exponential decay modulated with the rate of angular separation $2\omega_{\rm r}\Delta\ell$. We note that while in our experiment there are both $\Delta\ell$-separated and $2\Delta\ell$-separated coherences, the fringe contrast of the average $\ket{D}$ state population is sensitive only to the $\Delta\ell$-separated coherences, and is insensitive $2\Delta\ell$-separated coherences. The experiment takes place in a regime where decoherence occurs well before the completion of the first oscillation at time $\tau=\pi/\omega_{\rm r}\Delta\ell$, so that
\begin{equation}
    C(\tau) = C_0\,\exp \left [-(\gamma \tau)^3\right ],
    \label{eq:Ct-highD}
\end{equation}
with
\begin{equation}
        \gamma = \left(\frac{\omega_{\rm r}^2\Delta\ell^2D}{3\hbar^2}\right)^{1/3}.
        \label{eq:gamma}
\end{equation}
$C_0$ less than unity accounts for imperfect operations. Fitting the measurements to this function yields the decoherence rate $\gamma$, see Fig.~\ref{fig:decoherence-measurements}(b). In Fig.~\ref{fig:decoherence-measurements}(c), we present triplets of decoherence rates for superpositions $\Delta\ell=1, 2, 3$ across various strengths of the system-environment interaction, which are quantified by the angular momentum diffusion coefficient. By varying the amplitude of injected noise, we tune the diffusion coefficient from 2.1\,$\hbar^2$/ms (ambient environment, no noise injected) to 1000\,$\hbar^2$/ms. The solid curves are derived directly from the solution to the master equation with no free fitting parameters.

Our experiment confirms three non-trivial predictions: (i) The shape of the measured contrast decay profile in Fig.~\ref{fig:decoherence-measurements}(b) matches the predicted profile \eqref{eq:Ct-highD}. In the picture of superposed time-dependent orientations, the $\tau^3$-dependence may be understood as the integral of an instantaneous decoherence rate that is quadratic in the angular separation. This is a signature of the sine-squared orientational scaling of rotational decoherence in the small-angle limit, since for small angular separation one obtains $\sin^2[\Delta \phi(t)] = \sin^2(2\omega_{\rm r}\Delta \ell t) \approx (2\omega_{\rm r}\Delta \ell t)^2$. (ii) The power-law scaling for the decoherence rate \eqref{eq:gamma} with $\Delta\ell$, $\gamma\sim\Delta\ell^{2/3}$, is confirmed by the fit in the inset of Fig.~\ref{fig:decoherence-measurements}(c). This yields $\gamma\sim\Delta\ell^{0.71(5)}$, which is consistent with the expected scaling. Repeating this analysis for the four diffusion coefficients in the main plot of Fig.~\ref{fig:decoherence-measurements}(c) and averaging the exponents, we find $\gamma\sim\Delta\ell^{0.65(4)}$. As with the shape of the coherence profile in time, the scaling $\gamma \sim \Delta\ell^{2/3}$ also results directly from the sine-squared orientational scaling, since $\Delta\ell$ sets the speed of angular separation. (iii) The power-law scaling of the decoherence rate \eqref{eq:gamma} with diffusion coefficient, $\gamma\sim D^{1/3}$, can be seen in our measurements in Fig.~\ref{fig:decoherence-measurements}(c), where the slope of the solid curves follows this $1/3$ power law. We emphasize that the solid curves come directly from \eqref{eq:gamma} with no free parameters. Thus in addition to finding agreement with all expected scalings of the rotational decoherence process, we also find that the absolute values of the measured decoherence rates agree with predictions from the master equation. We see agreement over the full range of measured diffusion coefficients.

\emph{Discussion.}---In summary, we have presented an experimental study of the decoherence dynamics of a trapped-ion planar rotor due to a noisy electric field. We have verified the theoretical prediction that the decoherence rate of a superposition of two orientations is proportional to $\sin^2(\phi-\phi')$ in the small-angle limit. In addition, by independently engineering and measuring the rate of angular momentum diffusion, we have confirmed the dependence of rotational decoherence on angular momentum diffusion. These results serve as a controlled test of the theory of rotor decoherence, which constitutes a necessary step towards utilizing rotational quantum coherence for sensing, quantum information processing, and fundamental tests \cite{DeMille2002, Grimsmo2020, Albert2020, Zhou2011, Lechner2013, Sundar2018, Ahn2020, Stickler2018, Ma2020}. Our work may serve as a basis for the observation of decoherence-free subspaces in the rotational degree of freedom \cite{Pedernales2020}, for future studies of surface noise beyond heating measurements with single ions \cite{Martinetz2022}, and for first experiments creating massive superposition states with nanoscale spin rotors \cite{Rusconi2022}. Finally, quantitatively understanding rotor decoherence in planar surface traps is a crucial step first step towards probing the exchange phase of distant atoms in trapped rotor interferometer \cite{Roos2017}.

\vspace{5mm}
\emph{Acknowledgements.}---Funding provided by the NSF Grant No. PHY 2011973 and by AFOSR Grant No. FA9550-23-1-0546 and  some institutional support by NSF Grant No. 2016245. B.A.S. acknowledges support by the Deutsche Forschungsgemeinschaft (DFG, German Research Foundation) – 510794108 and by the Carl Zeiss foundation through the project QPhoton.

\appendix

\section*{End Matter}

{\it Experimental details---} The rf electrodes produce a three-dimensional harmonic potential centered 180\,\textmu m above the trap surface, whose confinement strength is characterized by secular trap frequencies $\omega_z={2\pi\times2.87}$\,MHz normal to the trap surface and $\omega_x=\omega_y={2\pi\times1.44}$\,MHz in the transverse plane. The 729\,nm laser addresses the full ion crystal nearly normal to the rotor, at a small angle of 5.6$\degree$ with respect to the normal of the rotor plane, in order to maximize coupling to the desired sideband orders \cite{Urban2019}. A uniform magnetic field of 409\,\textmu T separates the Zeeman sublevels of the electronic states to allow frequency selectivity of the sublevels of interest. State measurement is performed by applying cycling transition laser light at 397\,nm, which is resonant with a dipole-allowed transition from the $\ket{S}$ state but off-resonant from any transition from the $\ket{D}$ state. This allows state discrimination by counting scattered photons.

\section{Theoretical description}
\subsection{Derivation of the master equation}
We consider a planar rigid rotor formed by two ions with charges $e$ at positions $r_{1,2} = z {\bf e}_z \pm r {\bf e}_\rho(\phi)$, where $z$ is the height above the surface trap, $2r$ is the ion separation, and the angle $\phi$ describes the rotor orientation. The potential energy of the rotor in the quadrupole potential follows as
\begin{equation}
    V(\phi) = -e r^2 {\bf e}_\rho(\phi) \cdot \left [ \nabla \otimes {\bf E}(z{\bf e}_z,t)\right ] {\bf e}_\rho(\phi),
\end{equation}
where ${\bf E}({\bf r},t)$ is the local field strength at time $t$. The vector gradient $\nabla \otimes {\bf E}(z{\bf e}_z,t)$ is symmetric and traceless, and determined by the trap geometry. Using that $L_\pm = e^{\pm i \phi}$, this potential energy can be rewritten up to an additive constant as
\begin{equation}\label{randompotential-supp}
    V_t(\phi) = \frac{er^2}{4} \left (\varepsilon_t L_+^2 + \varepsilon_t^* L_-^2\right ),
\end{equation}
where we defined the complex-valued combination of field gradients
\begin{equation}\label{Seq:epsilont}
    \varepsilon_t = -\partial_x E_x +\partial_y E_y + 2i \partial_x E_y,
\end{equation}
whose magnitude is determined by the trapping geometry as well as the time-dependent voltage applied to the control electrodes.

The quantum state $\ket{\psi_t}$ of the rotor evolves according to the Schrödinger equation with the free Hamiltonian $H_0 = L_z^2/2I$ and the potential \eqref{randompotential-supp}. Transformation to the interaction picture $\ket{\psi_t} = \exp(-iH_0t/\hbar) \ket{\chi_t}$ shows that the rotor evolves according to the interaction picture potential $\widetilde{V}_t = \exp(i H_0 t/\hbar) V_t \exp(-iH_0 t/\hbar)$, which can be further evaluated by using that
\begin{equation}
    L_\pm^2(t) = e^{i H_0 t/\hbar} L_\pm^2 e^{-iH_0 t/\hbar} = e^{2i \hbar t/I} L_\pm^2 e^{\pm 2it L_z/I},
\end{equation}
as follows from the commutation relations $[L_z,L_\pm] = \pm \hbar L_\pm$ and $L_\pm L_\mp=\mathbb{I}$. The interaction picture Schrödinger equation is thus given by $i\hbar \partial_t \ket{\chi_t} = \widetilde{V}_t \ket{\chi_t}$ with
\begin{equation}\label{eq:potential}
    \widetilde{V}_t = \frac{e r^2}{4} \left ( e^{2i\hbar t/I} \varepsilon_t L_+^2 e^{2itL_z/I} + {\rm h.c.}\right ).
\end{equation}
We now integrate this Schrödinger equation over the time interval $\Delta t \gg 1/\omega_{\rm rot}$ and iterate the resulting integral equation twice to obtain
\begin{align}\label{eq:iterateschroed}
    \ket{\chi_{t + \Delta t}} = & \ket{\chi_t} - \frac{i}{\hbar} \int_t^{t+\Delta t} dt' \widetilde{V}_{t'}\ket{\chi_t} \nonumber \\
    & - \frac{1}{\hbar^2}\int_t^{t + \Delta t} dt' \int_t^{t'}dt'' \widetilde{V}_{t'} \widetilde{V}_{t''} \ket{\chi_{t''}}.
\end{align}
To express that $\ket{\chi_t}$ evolves little in $\Delta t$, we approximate $\ket{\chi_{t''}} \simeq \ket{\chi_t}$. 

This equation describes how the dynamics of the pure state $\ket{\chi_t}$ is influenced by the time-dependent voltage applied to the control electrodes. This voltage is chosen randomly, so that the rotor state is described by the density operator $\widetilde{\rho} = \mathbb{E}[\ket{\chi_t}\bra{\chi_t}]$, where $\mathbb{E}$ denotes the ensemble average over many repetitions of the experiment. The equation of motion for $\widetilde{\rho}$ can be derived by calculating 
\begin{equation}
    \Delta \widetilde{\rho} = \mathbb{E}(\ket{\chi_{t + \Delta t}}\bra{\chi_{t + \Delta t}}) - \ket{\chi_t}\bra{\chi_t}
\end{equation}
by inserting Eq.~\eqref{eq:iterateschroed}, which yields
\begin{align}\label{eq:deltarho}
    \Delta \widetilde{\rho} = & \frac{1}{\hbar^2} \int_t^{t + \Delta t} dt' \int_{t}^{t+\Delta t} dt'' \mathbb{E} \left (\widetilde{V}_{t'} \widetilde{\rho}\widetilde{V}_{t''}\right ) \notag \\
    & - \frac{1}{\hbar^2} \int_t^{t + \Delta t} dt'\int_t^{t'}dt'' \left \{ \mathbb{E}\left (\widetilde{V}_{t'} \widetilde{V}_{t''}\right ),\widetilde{\rho}\right \}.
\end{align}
Here, $\{A,B\} = AB + BA$ denotes the anticommutator.

The time integrals appearing in Eq.~\eqref{eq:deltarho} can be calculated for rapidly revolving rotor states $\ket{\chi_t}$, which are localized in angular momentum close to the quantum number $\overline{\ell} = I \omega_{\rm rot}/\hbar \gg 1$ with angular momentum spread $\sigma_\ell \ll \ell$. Since this implies $\exp(\pm 2 i t L_z/I)\ket{\chi_t} \simeq e^{\pm 2 i \omega_{\rm rot} t} \ket{\chi_t}$ one is left with the integrals
\begin{subequations}\label{eq:psdintegrals}
\begin{equation}\label{eq:integral}
\int_{t}^{t + \Delta t}dt'\int_{t}^{t + \Delta t}dt'' \mathbb{E}(\varepsilon_{t'}\varepsilon^*_{t''}) e^{2i\omega_{\rm rot}(t' - t'')} \simeq \Delta t S_\varepsilon(2\omega_{\rm rot})
\end{equation}    
and
\begin{equation}
    \int_{t}^{t + \Delta t}dt'\int_{t}^{t'}dt'' \mathbb{E}(\varepsilon_{t'}\varepsilon^*_{t''}) e^{2i\omega_{\rm rot}(t' - t'')} \simeq \frac{\Delta t}{2} S_\varepsilon(2\omega_{\rm rot})
\end{equation}
\end{subequations}
upon inserting (\ref{eq:potential}) and neglecting rapidly rotating terms $\exp[\pm i 2\omega_{\rm rot} (t' + t'')] \simeq 0$ in the rotating-wave approximation.
In both integrals we used that $\varepsilon_t$ is described by a real, stationary, and time-inversion-invariant stochastic process with zero mean, multiplied by a fixed complex phase factor (as determined by (\ref{Seq:epsilont})), to identify the power spectral density
\begin{equation}
S_\varepsilon(\omega) = \int_{-\infty}^\infty d\tau \mathbb{E}(\varepsilon_t \varepsilon^*_{t - \tau}) e^{i\omega \tau}.
\end{equation}
The statistical properties of the the applied voltage directly translate to those of $\varepsilon_t$ since retardation effects in the field propagation can be neglected at the short distances of the experiment. 
The time interval $\Delta t$ will be taken large compared to the decay time of the noise correlations. Moreover, the noise filter bandwidth of 20\,kHz is large in comparison to the width of the rotation frequencies $\hbar \sigma_\ell/I \simeq 1$\,kHz, justifying the approximation before Eq.~\eqref{eq:psdintegrals}.

Inserting the integrals \eqref{eq:psdintegrals} into  equation \eqref{eq:deltarho} shows that the ensemble-averaged quantum state follows the Markovian master equation
\begin{align}
    \partial_t \widetilde{\rho}  \simeq \frac{\Delta \widetilde{\rho}}{\Delta t} =  \frac{D}{4\hbar^2} \left ( L_+^2 \widetilde{\rho} L_-^2 + L_-^2 \widetilde{\rho} L_+^2 - 2 \widetilde{\rho}\right ),
\end{align}
with the diffusion coefficient $D$ given by $e^2r^4S_{\varepsilon}(2\omega_\text{rot})/4$ as stated in the main text. Transforming back to the Schrödinger picture yields (under the same assumptions as above) the master equation (2).

Note that the diffusion coefficient can also be obtained in a classical picture. Starting from the equation of motion in the noisy potential \eqref{randompotential-supp},
\begin{equation}\label{eq:eom}
    \dot{L}_z = - i\frac{e r^2}{2} \left (\varepsilon_t e^{2i\phi} - \varepsilon^*_t e^{-2i\phi} \right ),
\end{equation}
together with $\dot{\phi} = L_z/I$, it follows that $\mathbb{E}(L_z) = {\rm const}$. In order to obtain the diffusion coefficient, we first formally solve \eqref{eq:eom} for a time step $\Delta t$ in the co-rotating frame $\phi \to \phi + \omega_{\rm rot} t$,
\begin{align}
    L_z(t + \Delta t)  = & L_z(t) - i\frac{e r^2}{2} \int_{t}^{t+\Delta t} dt' \left (\varepsilon_{t'} e^{2i\phi + 2i \omega_{\rm rot}t'} \right. \notag \\
    &\left. - \varepsilon^*_{t'} e^{-2i\phi - 2i \omega_{\rm rot}t'} \right ).
\end{align}
Here we use that $\phi(t') \simeq \phi(t)$ under the same approximation as after \eqref{eq:deltarho}. A rotating wave approximation and \eqref{eq:integral} yield the second moment
\begin{align}
    \mathbb{E}&[L_z^2(t+\Delta t)] -\mathbb{E}[L_z^2(t)] \notag\\
    & \simeq \frac{e^2 r^4}{2} \int_t^{t+\Delta t} dt' \int_t^{t+\Delta t} dt'' \mathbb{E}(\varepsilon_{t'} \varepsilon^*_{t''}) e^{2i \omega_{\rm rot}(t' - t'')} \notag \\
    & \simeq 2 D \Delta t.
\end{align}
Drawing the limit $\Delta t \to 0$, we thus obtain $\partial_t \mathbb{E}(L_z^2) = 2D$ as also implied by the quantum master equation \eqref{eq:master-equation}. While angular momentum diffusion can be understood classically, the observed decoherence of rotational superpositions is a genuine signature of the quantum master equation \eqref{eq:master-equation}.

\subsection{Time evolution in the noise field}

The time evolution under the master equation (2) can be determined by using Eq.~\eqref{eq:iterateschroed} for a pure state $\ket{\chi_t}$ and then calculating the ensemble average over many repetitions of the experiment. Specifically, approximating $\ket{\chi_{t''}}$ on the right hand side by $\ket{\chi_t}$ (see above), the operator
\begin{align}
    W(t) = & \mathbb{I} - \frac{i}{\hbar} \int_t^{t+\Delta t} dt' \widetilde{V}_{t'} \nonumber \\
    & - \frac{1}{\hbar^2}\int_t^{t + \Delta t} dt' \int_t^{t'}dt'' \widetilde{V}_{t'} \widetilde{V}_{t''} 
\end{align}
propagates the state over the time interval $\Delta t$, $\ket{\chi_{t+\Delta t}} = W(t)\ket{\chi_t}$. Concatenating $N=t/\Delta t\gg1$ of these timesteps, we can express the Schrödinger-picture state at time $t$ as 
\begin{equation}\label{eq:timeevolv}
    \ket{\psi_t} = e^{-iH_0 t/\hbar} U_t \ket{\psi_0},
\end{equation}
where $\ket{\psi_0}$ is the initial state and
\begin{equation}\label{eq:unitarydecomp}
U_t \simeq W\left ( t - \Delta t \right )\cdots W\left ( \Delta t \right )W\left ( 0 \right ),
\end{equation}
When calculating expectation values in the ensemble average, we can use that the random fields $\varepsilon_t$ are uncorrelated between different time intervals $[n\Delta t, (n+1)\Delta t]$. One can thus decompose the ensemble average over the random noise trajectory $\mathbb{E} ( \cdot )$ by the product of ensemble averages $\mathbb{E}_n(\cdot)$ at fixed times $t_n = n\Delta t$, so that
\begin{equation}\label{eq:ensembleaver}
    \mathbb{E}(\cdot) = \lim_{N \to \infty} \mathbb{E}_{N-1}(\mathbb{E}_{N-2}( \cdots \mathbb{E}_0(\cdot))).
\end{equation}
This property will be used below to determine the outcome of the Ramsey interference protocol.

The Ramsey scheme discussed below requires the time evolution \eqref{eq:timeevolv} of a coherent rotor state $\ket{\Psi_{\ell}}$ characterized by a mean angular momentum quantum number $\ell$ and width $\sigma_\ell$ with $\ell\gg\sigma_\ell\gg 1$. Using that $L_\pm^2(t)\ket{\Psi_\ell} \simeq e^{\pm 4i \ell\omega_{\rm r} t} \ket{\Psi_\ell}$, with $ \omega_{\rm r} = \hbar /2I$, yields
\begin{align}\label{eq:timeevol1}
    \ket{\psi_t} &\simeq u_\ell(t)  e^{- i H_0 t/\hbar} \ket{\Psi_\ell} \nonumber
    \\
    &\simeq u_\ell(t) e^{i \ell^2 \omega_{\rm r} t} e^{- i 2 \ell\omega_{\rm r} t L_z/\hbar} \ket{\Psi_\ell}.
\end{align}
Here, $u_\ell(t)$ is the c-number obtained by replacing in $U_t$ all operators $L_\pm^2(t_n)$ by $\exp( \pm 4i  \ell\omega_{\rm r} t_n)$. In the second expression, the unitary operator $\exp( - i 2 \ell\omega_{\rm r}  t L_z/\hbar)$ serves to displace the rotor state by the angle $2 \ell\omega_{\rm r}  t$, while  the global phase factor results from neglecting dispersion.

The Ramsey scheme also requires a generalization of \eqref{eq:timeevol1} for the the case of a coherent rotor state displaced by the angle $\varphi$. Using that
\begin{equation}
    L_\pm^2(t) e^{-i \varphi L_z/\hbar} = e^{\pm 2i\varphi}e^{- i \varphi L_z/\hbar} L_\pm^2(t),
\end{equation}
we obtain
\begin{align}\label{eq:timeevol2}
    \ket{\psi_{t}} &\simeq u_\ell(t|\varphi) e^{-iH_0 t/\hbar}e^{-i\varphi L_z/\hbar}\ket{\Psi_\ell}    
    \nonumber
    \\
    &\simeq u_\ell(t|\varphi) e^{i \ell^2 \omega_{\rm r} t} e^{-i(\varphi + 2 \ell\omega_{\rm r} t)L_z/\hbar}\ket{\Psi_\ell}
    .
\end{align}
Like above, the c-numbers $u_\ell(t|\varphi)$ are obtained by replacing in $U_t$ all operators $L_\pm^2(t_n)$ by $\exp[ \pm 2i (\varphi + 2 \ell\omega_{\rm r} t_n)]$.

\subsection{Rabi pulses}

This section derives the action of a Rabi pulse on the four internal levels and the rotational degrees of freedom of the two-ion rotor. We start with the light-matter Hamiltonian
\begin{equation}
    H_{\rm L} = \sum_{j = 1}^2 \sigma^+_j  f_{\rm L}({\bf r}_j,t) + {\rm h.c.},
\end{equation}
where $\sigma_j^+ = \ket{D}_j \bra{S}_j$ describes the excitation of the quadrupole transition of atom $j$, and $f_{\rm L}({\bf r}_j,t)$ is determined by the relevant field gradient at the position of the $j$-th atom. Assuming plane-wave illumination with inclination $\theta$ to the rotor plane, we have
\begin{equation}
f_{\rm L}({\bf r}_j,t) = \hbar g_0 e^{-i(k z\cos \theta  + \omega t)} e^{ i(-1)^j k r \sin \theta \cos \phi},
\end{equation}
where $g_0$ is the coupling rate, $\omega$ is the laser frequency, and $k = \omega/c$. Using that
\begin{equation}
    \bra{\ell + \Delta \ell}e^{\pm i k r \sin \theta \cos \phi}\ket{\ell} = (\pm i)^{\Delta \ell} J_{\Delta \ell}(kr \sin \theta),
\end{equation}
one obtains
\begin{align}\label{eq:lightmatterh}
    H_{\rm L} = & \hbar g_0 e^{-i(kz\cos\theta+\omega t)}  \sum_{j = 1}^2 \sum_{\Delta \ell\in \mathbb{Z}} (- 1)^{j\Delta \ell} i^{\Delta \ell} \notag \\
    & \times  J_{\Delta \ell}(kr\sin\theta)\sigma_j^+ L_+^{ \Delta \ell} + {\rm h.c.},
\end{align}

The light-matter Hamiltonian \eqref{eq:lightmatterh} can be further simplified by first transforming into the frame co-rotating with the laser field and then into the interaction picture with respect to $H_{\rm rot} = \hbar \Delta \omega
(\sigma_1^+\sigma_1^-+\sigma_2^+\sigma_2^-)+H_0$, where $\Delta\omega = \omega_0 -\omega$ denotes the laser detuning from the electronic transition frequency $\omega_0$. This yields
\begin{align}
    H_{\rm L}(t) = & e^{i H_{\rm rot}t/\hbar}H_{\rm L} e^{-iH_{\rm rot}t/\hbar} \notag \\
     =& \hbar g_0 e^{-ikz\cos\theta} \sum_{j = 1}^2 \sum_{\ell \in \mathbb{Z}} \sum_{\Delta \ell \in \mathbb{Z}} (-1)^{j\Delta \ell}i^{\Delta \ell} J_{\Delta \ell}(kr\sin \theta) \notag \\
      & \times e^{ it [\Delta\omega +  \Delta \ell (2 \ell + \Delta \ell)\omega_{\rm r}]} \sigma_j^+ \ket{\ell + \Delta \ell}\bra{\ell} + {\rm h.c.}
\end{align}
For rapidly rotating states, where $2\ell + \Delta \ell \simeq 2 \ell$, the laser frequency is tuned in resonance with the angular momentum transition $\Delta \ell$ by choosing $\omega \simeq \omega_0 + 2 \ell\omega_{\rm r} \Delta \ell$, so that
\begin{equation}
    H_{\rm L}(t) \simeq \hbar g \left [\sigma_1^+ L_+^{\Delta \ell} + (-1)^{\Delta \ell} \sigma_2^+ L_+^{\Delta \ell} \right ] + {\rm h.c.},
\end{equation}
where we defined
\begin{equation}
    g = g_0 e^{ -i k z \cos \theta} (-i)^{\Delta \ell}J_{\Delta \ell}(kr \sin\theta).
\end{equation}
The excitation of the atom's electronic state is thus accompanied by an angular momentum kick of strength $\Delta \ell$. Applying the laser field for a short time period $t$, as described by the unitary
\begin{align}
U_{\rm L}(t) = & \exp \left  [ - igt \left (\sigma_1^+ L_+^{\Delta \ell}+{\rm h.c.} \right ) \right ]    \notag \\
 & \times \exp \left  [ - igt (-1)^{\Delta \ell}  \left (\sigma_2^+ L_+^{\Delta \ell}+{\rm h.c.} \right ) \right ],   
\end{align}
gives rise to Rabi oscillations, which can be used to implement $\pi/2$-pulses and $\pi$-pulses. Specifically, choosing $gt  = -i\pi/4$ implements a $\pi/2$-pulse,
\begin{align}\label{eq:pi2pulse}
    X_{\pi/2} = & \frac{\mathbb{I}}{2} - \frac{\sigma_1^+ + (-1)^{\Delta \ell} \sigma_2^+}{2} L_+^{\Delta \ell} +  \frac{\sigma_1^- + (-1)^{\Delta \ell} \sigma_2^-}{2} L_-^{\Delta \ell}\notag\\
    & + \frac{(-1)^{\Delta \ell} }{2} \left (\sigma_1^+\sigma_2^+ L_+^{2\Delta \ell} + \sigma_1^-\sigma_2^- L_-^{2\Delta \ell} \right. \notag \\
    & \left. - \sigma_1^+\sigma_2^- - \sigma_1^-\sigma_2^+\right ).
\end{align}
Likewise, choosing $gt  =-i\pi/2$ gives the $\pi$-pulse
\begin{align}\label{eq:pipulse}
    X_\pi = & (-1)^{\Delta \ell} \left (\sigma_1^+\sigma_2^+ L_+^{2\Delta \ell} + \sigma_1^-\sigma_2^- L_-^{2\Delta \ell} \right. \nonumber \\
    & \left.- \sigma_1^+\sigma_2^- - \sigma_1^-\sigma_2^+ \right ) .
\end{align}

\subsection{Ramsey scheme}

The interferometer sequence, as sketched in Fig.~\ref{fig:intseqsketch}, consists of the following steps:
\begin{figure}
    \includegraphics[width = 0.49\textwidth]{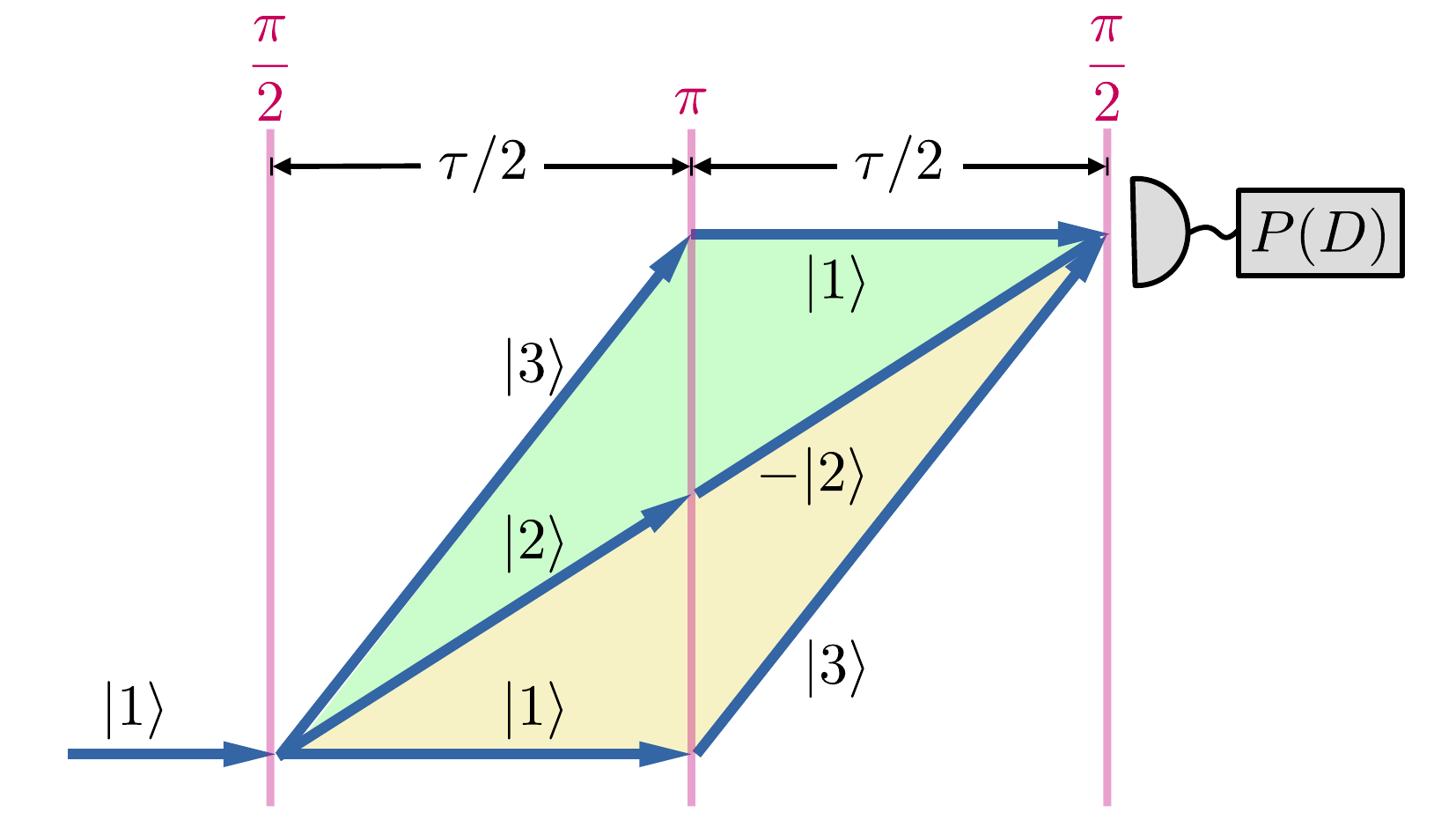}
    \caption{Sketch of the rotational interferometer sequence consisting of the two $\pi/2$-pulses and an intermediate $\pi$-pulse to implement a Hahn echo at separation $\tau/2$. The three involved two-ion rotor states, defined in Eqs.~\eqref{eq:threestates}, revolve with mean angular momentum $\overline{\ell}$, $\overline{\ell}+\Delta \ell$, and $\overline{\ell} + 2\Delta \ell$, respectively. The shaded regions indicate the coherences between superposition branches as read-out by the measurement of the mean excitation probability $P(D)$, see Eq.~\eqref{eq:pd}.}\label{fig:intseqsketch}
\end{figure}

\begin{enumerate}[leftmargin=*,itemsep = 0.2mm]
    \item[(0)] {\bf Initialization---}The rotor is prepared in the coherent rotor state $\ket{\Phi_0} = \ket{SS,\Psi_{\overline{\ell}}}$ with mean angular momentum $\overline{\ell} \gg \sigma_\ell$.
    
    \item[(1)] {\bf Creation of the superposition---}At $t = 0$, a $\pi/2$-pulse creates a superposition of angular momentum states $\ket{\Phi_1} = X_{\pi/2}\ket{\Phi_0}$. Direct application of Eq.~\eqref{eq:pi2pulse} yields the threefold superposition
    \begin{align}\label{eq:phi1}
        \ket{\Phi_1} = & \frac{1}{2}\left ( \ket{1} - \sqrt{2}\ket{2} + \ket{3}\right ),
    \end{align}
    with the three orthonormal states
    \begin{subequations}\label{eq:threestates}
    \begin{align}
        \ket{1} = & \ket{SS,\Psi_{\overline{\ell}}}\\
        \ket{2} = & \frac{1}{\sqrt{2}}\left ( \ket{DS} + (-1)^{\Delta \ell} \ket{SD} \right )\ket{\Psi_{\overline{\ell}+\Delta \ell}}\\
        \ket{3} = & (-1)^{\Delta \ell} \ket{DD,\Psi_{\overline{\ell}+2\Delta \ell}}
    \end{align}
    \end{subequations}
    The laser pulse thus entangles the two ion's internal state with that of the rotational degree of freedom. Whether $\Delta \ell$ is even or odd determines whether the $\pi/2$ pulse creates the even or odd superposition of $\ket{DS}$ and $\ket{SD}$.

    \item[(2)] {\bf Time evolution---}This is followed by a time evolution according to Eq.~\eqref{eq:timeevolv} in presence of the environment until $t = \tau/2$, $\ket{\Phi_2} = e^{-i H_{\rm rot} \tau/2\hbar}U_{\tau/2} \ket{\Phi_1}$. Using the result \eqref{eq:timeevol1} yields
    \begin{align}
        \ket{\Phi_2} = & \frac{1}{2}\left [ a_{\overline{\ell}}T_{\overline{\ell}} \ket{1} - \sqrt{2} e^{-i\Delta\omega \tau/2}a_{\overline{\ell} + \Delta \ell}T_{\overline{\ell} + \Delta \ell}\ket{2} \right. \nonumber\\
        & \left. \vphantom{a_{\overline{\ell}}T_{\overline{\ell}} \ket{1} - \sqrt{2} a_{\overline{\ell} + \Delta \ell}T_{\overline{\ell} + \Delta \ell}\ket{2}}+ e^{-i\Delta\omega \tau}a_{\overline{\ell} + 2\Delta \ell}T_{\overline{\ell} + 2\Delta \ell}\ket{3}\right ].
    \end{align}
    Here, we introduced the weights
    \begin{equation}
    a_\ell = e^{i  \ell^2 \omega_{\rm r}\tau/2} u_{\ell}\left ( \frac{\tau}{2}\right ) 
    \end{equation}
    and the angular displacement operators
    \begin{equation}
         T_\ell =  \exp\left ( - i \frac{\ell \omega_{\rm r}\tau L_z}{\hbar}\right )
    \end{equation}
    
    \item[(3)] {\bf Hahn echo pulse---}At $t = \tau/2$, a $\pi$-pulse implements a Hahn echo $\ket{\Phi_3} = X_{\pi}\ket{\Phi_2}$. Using Eq.~\eqref{eq:pipulse} this becomes
    \begin{align}
        \ket{\Phi_3} = & \frac{1}{2}\left [ a_{\overline{\ell}}T_{\overline{\ell}}\ket{3} + e^{-i\Delta\omega \tau/2}\sqrt{2} a_{\overline{\ell} + \Delta \ell}T_{\overline{\ell} + \Delta \ell}\ket{2} \right. \nonumber \\
        & \left.\vphantom{a_{\overline{\ell}}T_{\overline{\ell}} \ket{1} - \sqrt{2} a_{\overline{\ell} + \Delta \ell}T_{\overline{\ell} + \Delta \ell}\ket{2}} + e^{-i\Delta\omega \tau} a_{\overline{\ell} + 2\Delta \ell}T_{\overline{\ell} + 2\Delta \ell}\ket{1}\right ].
    \end{align}
    
    \item[(4)] {\bf Time evolution---}This is followed by the second time evolution in presence of the environment until $t = \tau$, $\ket{\Phi_4} = e^{-i H_{\rm rot} \tau/2\hbar}U_{\tau/2} \ket{\Phi_3}$. Using Eq.~\eqref{eq:timeevol2} yields
    \begin{align}\label{eq:phi4}
        \ket{\Phi_4} = & \frac{1}{2} e^{-i\Delta\omega \tau}T_{\overline{\ell} + \Delta\ell}^2 \left [b_{\overline{\ell} + 2\Delta \ell}^{\overline{\ell}} a_{\overline{\ell}} \ket{3} + \sqrt{2} b_{\overline{\ell}+\Delta \ell}^{\overline{\ell}+\Delta \ell} a_{\overline{\ell} + \Delta \ell}\ket{2} \right. \nonumber \\
        & \left. + b_{\overline{\ell}}^{\overline{\ell}+2\Delta \ell}a_{\overline{\ell} + 2\Delta \ell}\ket{1}\right ],
    \end{align}
    where we abbreviated the weights
    \begin{equation}
        b_\ell^{\ell'} = e^{i \ell^2\omega_{\rm r}\tau/2} u_\ell \left ( \frac{\tau}{2} \Big \vert \omega_{\rm r}\tau \ell'\right ).
    \end{equation}
    Note that all states in the superposition \eqref{eq:phi4} are translated by the same angle $2(\overline{\ell}+\Delta \ell)\omega_{\rm r}\tau$, meaning that the interferometer can now be closed.
    
    \item[(5)] {\bf Closing of the interferometer---}A second $\pi/2$-pulse $\ket{\Phi_5} = X_{\pi/2}\ket{\Phi_4}$ yields
    \begin{align}\label{eq:phi5}
        \ket{\Phi_5} = &\frac{1}{4} e^{-i\Delta\omega \tau} T^2_{\overline{\ell}+\Delta \ell}\left [ \sqrt{2} \left(b_{\overline{\ell}+2\Delta\ell}^{\overline{\ell}} a_{\overline{\ell}} - b^{\overline{\ell}+2\Delta\ell}_{\overline{\ell}} a_{\overline{\ell}+2\Delta \ell} \right ) \ket{2}\right. \nonumber\\
        & + \left (b_{\overline{\ell}+2\Delta \ell}^{\overline{\ell}} a_{\overline{\ell}} + 2 b_{\overline{\ell}+\Delta \ell}^{\overline{\ell}+\Delta \ell} a_{\overline{\ell} + \Delta \ell } + b^{\overline{\ell}+2\Delta \ell}_{\overline{\ell}} a_{\overline{\ell}+2\Delta \ell}\right )\ket{1} \nonumber \\
        & \left. + \left (b_{\overline{\ell}+2\Delta \ell}^{\overline{\ell}} a_{\overline{\ell}} - 2 b_{\overline{\ell}+\Delta \ell}^{\overline{\ell}+\Delta \ell} a_{\overline{\ell} + \Delta \ell } + b^{\overline{\ell}+2\Delta \ell}_{\overline{\ell}} a_{\overline{\ell}+2\Delta \ell}\right )\ket{3}\right ]
    \end{align}
    
    \item[(6)] {\bf Measurement---}Measuring the mean flourescence light of the two atoms yields the mean probability of being in the excited state,
    \begin{align}
        P(D) =&  {\rm Prob}(DD) + \frac{1}{2}\left[ {\rm Prob}(DS) + {\rm Prob}(SD) \right] \nonumber\\
        = & \frac{1}{2} + \frac{1}{2} \left [ {\rm Prob}(DD) - {\rm Prob}(SS)\right ].
    \end{align}
Using the final state \eqref{eq:phi5} and averaging over many repetitions of the experiment yields
    \begin{align}\label{eq:pd}
        P(D) =& \frac{1}{2} - \frac{1}{4} {\rm Re}\left [ \mathbb{E}\left (b_{\overline{\ell}+\Delta \ell}^{\overline{\ell}+\Delta \ell} (b_{\overline{\ell}+2\Delta \ell}^{\overline{\ell}})^* \right )\mathbb{E}(a_{\overline{\ell}+\Delta \ell} a^*_{\overline{\ell}})\right.\nonumber\\
        & \left. + \mathbb{E}\left (b_{\overline{\ell}+\Delta \ell}^{\overline{\ell}+\Delta \ell} (b^{\overline{\ell}+2\Delta \ell}_{\overline{\ell}})^*\right )\mathbb{E}(a_{\overline{\ell}+\Delta \ell} a^*_{\overline{\ell}+2\Delta \ell})\right ].
    \end{align}
    Here, we used that the coefficients $a_\ell$ and $b_\ell^{\ell'}$ are statistically uncorrelated as discussed in the context of Eq.~\eqref{eq:ensembleaver}. Next, we will explicitly calculate the ensemble averages in Eq.~\eqref{eq:pd}.
\end{enumerate}

\subsection{Ensemble averaged signal}

We are now in the position to evaluate the ensemble-averaged signal \eqref{eq:pd} by using the decomposition \eqref{eq:unitarydecomp} and the rule \eqref{eq:ensembleaver}. We start with
\begin{align}
    \mathbb{E}(a_{\overline{\ell}+ \Delta \ell} a^*_{\overline{\ell}}) = & \exp \left [ i \frac{\omega_{\rm r} \tau}{2} \Delta \ell(2\overline{\ell}+\Delta \ell) \right ]\nonumber\\
    &\times \mathbb{E}\left [u_{\overline{\ell}+\Delta \ell}\left ( \frac{\tau}{2}\right )u_{\overline{\ell}}^*\left ( \frac{\tau}{2}\right )\right].
\end{align}
The expectation value in the second line can be calculated by writing
\begin{align}\label{eq:expvalu}
    \mathbb{E}\left [u_{\overline{\ell}+\Delta \ell}\left ( \frac{\tau}{2}\right )u_{\overline{\ell}}^*\left ( \frac{\tau}{2}\right )\right] = \prod_{n,n' = 0}^N \mathbb{E}\left[w_{\overline{\ell}+\Delta \ell}(n\Delta t) w_{\overline{\ell}}^*(n'\Delta t)\right ],
\end{align}
where $\Delta t = \tau/2N$ with $N\gg1$, and we defined
\begin{align}\label{eq:wl}
    w_\ell(n\Delta t) = & 1 - \frac{i}{\hbar}\int_{n\Delta t}^{(n+1)\Delta t} dt' v_{\ell}(t') \nonumber \\
    & - \frac{1}{\hbar^2} \int_{n\Delta t}^{(n+1)\Delta t}dt' \int_{n\Delta t}^{t'} dt'' v_\ell(t') v_\ell(t''),
\end{align}
with
\begin{equation}
    v_\ell(t) = \frac{er^2}{4} ( \varepsilon_t  e^{4 i \ell \omega_{\rm r}t} + {\rm c.c.}).
\end{equation}
This latter function is obtained by replacing $L_\pm^2(t)$ in the potential energy \eqref{eq:potential} by the c-numbers $\exp(\pm 4 i \ell \omega_{\rm r}t)$.

Inserting Eq.~\eqref{eq:wl} into Eq.~\eqref{eq:expvalu} and using the same steps as in the derivation of the master equation yields
\begin{align}
    \mathbb{E} & \left [u_{\overline{\ell}+\Delta \ell}\left ( \frac{\tau}{2}\right )u_{\overline{\ell}}^*\left ( \frac{\tau}{2}\right )\right] \nonumber \\
    & \simeq  \prod_{n = 0}^N \left [1 - \frac{D \Delta t}{\hbar^2} \sin^2 \left [ \Delta \ell \left ( 2n + 1\right ) \omega_{\rm r} \Delta t \right ] \right ]\nonumber\\
    & \simeq  \prod_{n = 0}^N \exp\left [- \frac{D \Delta t}{\hbar^2} \sin^2 \left [ \Delta \ell \left ( 2n + 1\right )\omega_{\rm r}\Delta t \right ] \right ].
\end{align}
For $N\gg1$ the sum in the exponent can be approximated  by an integral, leading to
\begin{align}
    \mathbb{E}(a_{\overline{\ell}+ \Delta \ell} a^*_{\overline{\ell}}) \simeq &  \exp \left ( -\frac{D \tau}{4\hbar^2} \left [ 1-{\rm sinc} \left ( 2 \Delta \ell \omega_{\rm r} \tau \right ) \right ] \right ) \nonumber \\
    & \times \exp \left [ i \frac{\omega_{\rm r}\tau}{2} \Delta \ell(2\overline{\ell}+\Delta \ell) \right ].
\end{align}

In a similar fashion, one can evaluate the expectation value
\begin{align}
    \mathbb{E}(a_{\overline{\ell}+ \Delta \ell} a^*_{\overline{\ell} + 2\Delta \ell}) \simeq & \exp \left ( -\frac{D \tau}{4\hbar^2} \left [ 1-{\rm sinc} \left ( 2\Delta \ell\omega_{\rm r} \tau \right ) \right ] \right ) \nonumber\\
    & \times \exp \left [- i \frac{\omega_{\rm r} \tau}{2} \Delta \ell(2\overline{\ell}+3\Delta \ell) \right ]
\end{align}
It contains the same decoherence-induced decay as the expression above since only the relative angular momentum between the two superposition branches matters.

The expectation values containing $b_\ell^{\ell'}$ coefficients can be calculated by a straight-forward generalization replacing $v_\ell(t)$ by
\begin{equation}
    v_{\ell}^{\ell'}(t) = \frac{er^2}{4} ( \varepsilon_t  e^{i2 \ell' \omega_{\rm r}\tau} e^{4 i \ell \omega_{\rm r}t} + {\rm c.c.}).
\end{equation}
This yields
\begin{align}
    \mathbb{E}\left [ b_{\overline{\ell} + \Delta \ell}^{\overline{\ell} + \Delta \ell} (b^{\overline{\ell}}_{\overline{\ell} + 2\Delta \ell} )^*\right ] \simeq & \exp \left ( -\frac{D \tau}{4\hbar^2} \left [ 1-{\rm sinc} \left ( 2\Delta \ell\omega_{\rm r} \tau \right ) \right ] \right ) \nonumber\\
    & \times \exp \left [ -i \frac{\omega_{\rm r} \tau}{2} \Delta \ell(2\overline{\ell}+3\Delta \ell) \right ]
\end{align}
and 
\begin{align}
    \mathbb{E}\left [ b_{\overline{\ell} + \Delta \ell}^{\overline{\ell} + \Delta \ell} (b_{\overline{\ell}}^{\overline{\ell} + 2\Delta \ell} )^*\right ] \simeq & \exp \left ( -\frac{D \tau}{4\hbar^2} \left [ 1-{\rm sinc} \left ( 2 \Delta \ell \omega_{\rm r}\tau \right ) \right ] \right ) \nonumber\\
    & \times \exp \left [ i \frac{\omega_{\rm r} \tau}{2} \Delta \ell(2\overline{\ell}+\Delta \ell) \right ].
\end{align}

Putting everything together, we obtain the measurement signal
\begin{align}
P(D) = &  \frac{1}{2} - \frac{1}{2} \cos \left ( \omega_{\rm r} \tau \Delta \ell^2\right ) C(\tau),
\label{eq:pd-full}
\end{align}
with the reduction factor of the Ramsey fringe contrast due to rotational decoherence
\begin{equation}
    C(\tau) = \exp \left ( -\frac{D \tau}{2\hbar^2} \left [ 1-{\rm sinc} \left (2\Delta \ell\omega_{\rm r} \tau \right ) \right ] \right ).
\end{equation}

\section{Coherent effects on the contrast decay profile} \label{appendix:4-level-sys}

From \eqref{eq:pd-full} we see that the phase contrast, which decays from decoherence as $C(\tau)$ given by (4) in the main text, is additionally modulated by a factor
\begin{equation}
    C_\text{mod}(\tau) = \cos(\omega_\text{r}\Delta\ell^2\tau).
    \label{eq:Cmod}
\end{equation}
This arises from interference between the two relevant coherences of the rotor state: $\ket{\ell}\leftrightarrow\ket{\ell+\Delta\ell}$ and $\ket{\ell+\Delta\ell}\leftrightarrow\ket{\ell+2\Delta\ell}$, whose respective transition frequencies differ by $2\Delta\ell^2\omega_{\rm r} = 2\pi\times26$\,Hz\;$\times\,\Delta\ell^2$. This modulation is necessarily present in the coherence measurements presented in this work, which otherwise yield a contrast profile of the form $\exp[-(\gamma\tau)^3]$. 
Equation~\eqref{eq:Cmod} predicts a node in the fringe contrast at $\tau = \pi/(2\Delta\ell^2\omega_{\rm r})$. The modulation may therefore be safely neglected only if the coherence time is much shorter than the node time, $1/\gamma \ll \pi/(2\Delta\ell^2\omega_{\rm r})$. Otherwise, the effect of the modulation may be measurable before decoherence has occurred, and thus this effect must be accounted for in order to accurately estimate $\gamma$ from the measurements. We find from numerical simulations that the modulation profile may be appreciably influenced by imperfect operations, where only a partial initial coherence is created, which we therefore must also take into account.

\begin{figure}
    \centering
    \includegraphics[width=\columnwidth]{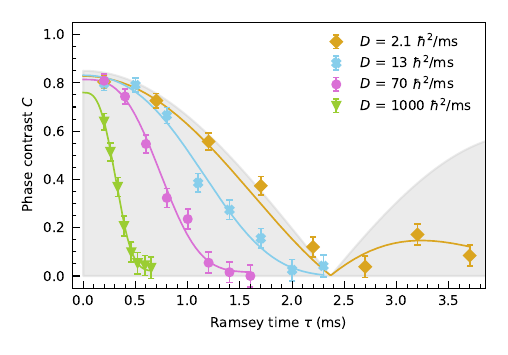}
    \caption{Coherent oscillations of phase contrast amplitude. All contrast decay measurements for $\Delta\ell=3$ measurements of decoherence presented in Fig.~\ref{fig:decoherence-measurements}(b) are shown here. The shaded region is the contrast profile of the coherent effect discussed in this section. Solid lines are fits of the form (5) from the main text, additionally multiplied by the shaded profile only if $\Delta\ell^2\omega_{\rm r}/2\pi\gamma > 0.1$. In these cases, the coherence time is sufficiently long that the effect is appreciable. Here, this applies to only the $D=2.1\,\hbar^2$/ms and $D=13\,\hbar^2$/ms curves.}
    \label{fig:contrast-oscillations}
\end{figure}

Figure~\ref{fig:contrast-oscillations} demonstrates this effect. Measured contrast decay curves for $\Delta\ell=3$ superpositions are shown, whose fits for $\gamma$ are shown in Fig.~\ref{fig:decoherence-measurements}(b). The shaded region shows the profile $C_\text{mod}(\tau)$, modified from \eqref{eq:Cmod} to account for imperfect operations (sub-unity initial contrast) such that it yields agreement with numerical simulations. In particular, \eqref{eq:Cmod} predicts a node in the contrast at $\tau = \pi/(2\Delta\ell^2\omega_r) = 2.1$\,ms, while the numerically simulated profile corrected for imperfect operations predicts this node to instead be at $t=2.4$\,ms, as shown in Fig~\ref{fig:contrast-oscillations}. If the coherence time is comparable or greater than this node time, then the measured decay profile differs significantly from (5) from the main text, and must be accounted for when extracting a decoherence rate. Including corrections due to imperfect operations, the contrast node is expected to occur at $\tau=4.8$\,ms for our $\Delta\ell=2$ measurements, and at $\tau=19$\,ms for our $\Delta\ell=1$ measurements. For most of our measurements, this effect is negligible. In estimating the decoherence rate $\gamma$ in Fig.~\ref{fig:decoherence-measurements}(b), we explicitly account for this effect only when $\Delta\ell^2\omega_{\rm r}/2\pi\gamma > 0.1$.

\section{Non-rigid effects} \label{appendix:non-rigid}

In the main text, we approximate the ion crystal as a rigid rotor. The leading-order non-rigid effect is centrifugal distortion: the moment of inertia increases as the square of the rotation frequency due to centrifugal force, effectively reducing the rotational constant $\omega_{\rm r}$ with increasing angular momentum. For an in-plane confinement strength such that the center-of-mass vibrational frequency is $\omega_x$, the fractional change to $\omega_{\rm r}$ due to finite rotation frequency $\omega_\text{rot}$ is given by $2\omega_\text{rot}^2/\omega_x^2 = 2.0\times10^{-2}$ for parameters used in this work. This quantity is smaller than the fractional uncertainty of any measurement presented, so we neglect this correction.

\section{Measurements of rotational friction} \label{appendix:friction}
\begin{figure}
    \centering
    \includegraphics[width=\columnwidth]{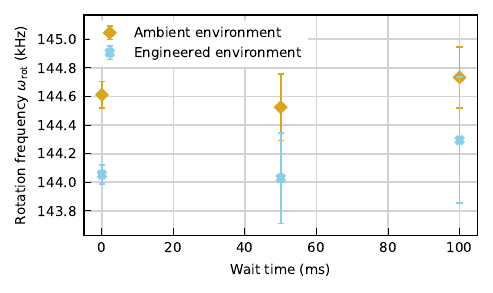}
    \caption{Measurements of rotational friction. The rotation frequency is measured as a function of wait time after preparation of a rapidly rotating state, both without without ($D = 19\,\hbar^2$/ms) and with ($D = 110\,\hbar^2$/ms) voltage noise injection. Over 100\,ms, no measurable slowdown is observed.}
    \label{fig:friction}
\end{figure}

A finite-temperature environment interacting with a quantum rotor will have a frictional effect in addition to angular momentum diffusion, which in general may affect the decoherence dynamics \cite{Stickler2018b}. We thus verify that the environment is well-approximated as infinite temperature by confirming that rotational friction is negligible. We measure rotational friction by recording the center frequency of a rotational sideband at various wait times after rotational state preparation. For increased sensitivity, the third rotational sideband frequency $\omega_{sb}^{(\Delta\ell=3)}$ is chosen, and the rotation frequency inferred as $\omega_\text{rot} = \omega_{sb}^{(\Delta\ell=3)}/3$. As shown in Fig.~\ref{fig:friction}, we do this for both an ambient environment ($D = 19\,\hbar^2$/ms) and an engineered environment ($D = 110\,\hbar^2$/ms). In both cases, we find the rate of slowdown to be consistent with zero. We bound the rate of slowdown (2$\sigma$ uncertainty of the slope of the fit) at $< 0.2\,\hbar$/ms in the case of the ambient environment, and $< 0.3\,\hbar$/ms in the case of the engineered environment. We thus neglect rotational friction in the theoretical treatment in the main text.

\vspace{10mm}
\section{Additional decoherence sources} \label{appendix:other-decoherence}

Two other potentially  relevant effects are decoherence of the ions' electronic state and changes of the rotor's moment of inertia due to fluctuations of the trap frequency. We rule out both of these as significant contributions.

\begin{figure}[b]
    \centering
    \includegraphics[width=\columnwidth]{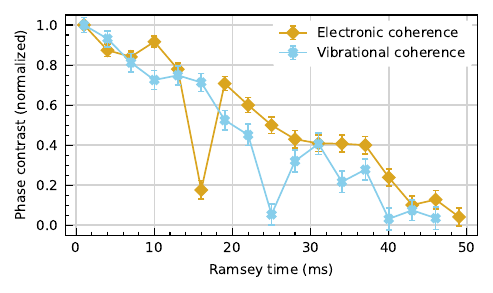}
    \caption{Measurements of other coherences. The electronic coherence measurement measures the coherence of the superposition $\ket{S}+\ket{D}$ of a single ion, and the motional coherence measurement measures the coherence of horizontal vibrational motion of a single ion, placing a bound on the stability of the rotor's moment of inertia. A Hahn echo pulse is included in both of these measurements to match the conditions of the rotational decoherence measurements.}
    \label{fig:other-coherences}
\end{figure}

Since the Ramsey experiment entangles the angular momentum state of the rotor with the electronic state of the ions, decoherence of the electronic states will also result in a loss of measured phase contrast. To measure the electronic coherence time, we trap a single ion and prepare it in the superposition $\ket{S}+\ket{D}$ with a Ramsey experiment including a Hahn echo pulse. This measurement is shown in Fig.~\ref{fig:other-coherences}, where we find the $1/e$ coherence time to be 38\,ms, much longer than any coherence measurement presented in this work.

Changes in  the rotor's moment of inertia conserve its angular momentum but change the rotation frequency, thereby also decohering a superposition of angular displacements. The stability of the moment of inertia $I$ is determined by the stability of the transverse trap frequency $\omega_x$, related by $I\propto\omega_x^{-4/3}$. We measure the stability of the trap frequency by trapping a single ion and preparing it in the Fock-state superposition $\ket{0} +\ket{1}$ with a Ramsey experiment including a Hahn echo pulse. We find that this vibrational motion has a $1/e$ coherence time of 32 ms, as shown in Fig.~\ref{fig:other-coherences}. When propagated to the case of angular momentum superpositions in the rotor for $\omega_\text{rot}=145$\,kHz, this yields an inferred rotational coherence time of $240$\,ms$/\Delta\ell$, again much longer than the actual rotational coherence times measured.

\end{document}